# Synthesis of Three-Dimensionally Interconnected Hexagonal Boron Nitride Networked Cu-Ni Composite

Zahid Hussain, Hye-Won Yang, Byung-Sang Choi

*Abstract*

A three-dimensionally interconnected hexagonal boron nitride (3Di-hBN) networked Cu-Ni (3Di-hBN-Cu-Ni) composite was successfully synthesized in situ using a simple two-step process which involved the compaction of mixed Cu-Ni powders (70 wt.% Cu and 30 wt.% Ni) into a disc followed by metal-organic chemical vapor deposition (MOCVD) process at 1000°C. During MOCVD, the Cu-Ni alloy grains acted as a template for the growth of hexagonal boron nitride (hBN) while decaborane and ammonia were used as precursors for boron and nitrogen, respectively. Boron and nitrogen atoms diffused into the Cu-Ni solution during the MOCVD process and precipitated out along the Cu-Ni interfaces upon cooling resulting in the formation of 3Di hBN-Cu-Ni composite. Energy-dispersive spectroscopic analysis confirmed the presence of boron and nitrogen atoms at the interfaces of Cu-Ni alloy grains. Optical microscopy examination indicated that there was a minimum amount of bulk hBN at certain compaction pressure (280 MPa) and sintering time (30 min). Scanning electron microscopy and transmission electron microscopy revealed that the interconnected network of hBN layers surrounding the Cu-Ni grains was developed in the 3Di-hBN-Cu-Ni composite. This 3Di-hBN network is expected to enhance the mechanical, thermal, and chemical properties of the 3Di-hBN-Cu-Ni composite. Moreover, foam-like 3Di-hBN extracted from 3Di-hBN-Cu-Ni composite could have further applications in the fields of biomedicine and energy storage.

# 1 Introduction

The development of two-dimensional (2D) materials has opened up the possibilities for their application in improving the properties of metals and alloys [1-8]. This is because 2D materials have the potential to alter the properties of metals at the nanoscale. A single layer of hexagonal boron nitride (hBN) is structurally similar to graphene (carbon system) where the hexagonal lattices are occupied by boron and nitrogen atoms. hBN has a lattice parameter of 25 nm and possesses extraordinary properties, such as high chemical stability [9], high mechanical strength [9], low density [10], high thermal stability [11], and high thermal shock resistance [12]. These excellent properties can be utilized to improve the performance of various metal matrix composites (MMCs) through the in-situ construction of three-dimensionally interconnected (3Di) hBN layers in their grain boundaries. Similar approaches have been employed by other researchers, who used 3D- networked graphene to tailor the properties of MMCs [8, 13-15]. For instance, Chen et al. [13] enhanced the yield and tensile strengths of copper by wrapping graphene around copper grains using chemical vapor deposition (CVD). They reported that the graphene acted as a barrier for dislocation movement, and consequently, the elastic modulus and strength were improved. Li et al. [8] reported that Cu-graphene composites had a higher thermal conductivity than pure copper because graphene offered an effective path for heat transfer between the Cu grain boundaries. Other properties, such as corrosion resistance and wear resistance, have also been improved [16, 17].

Because of its structure similarity with graphene, the hBN introduced to metal matrices can also impart similar effects. Several researchers have used boron nitride nanoparticles to enhance the strength, hardness, wear, and corrosion resistance of metallic alloys [18-22]. For instance, the microstructure and properties of BN/Ni-Cu composites fabricated by powder technology were reported by Tantaway et al. [23]. They found that the BN content led to a decrease in density and an increase in the hardness, electrical resistivity, and saturation magnetization of the composite. Omayma et al. [10] fabricated Cu/hBN nanocomposites by the PM route, in which powder mixtures of Cu and hBN were compacted and sintered at various temperatures ranging from 950°C to 1000°C. They found that the physical, mechanical and tribological properties of the composite were influenced by the hBN. However, we have not found any published studies in which metal matrices were reinforced by a 3Di network of hBN layers.

Recently, various techniques have been utilized to fabricate reinforced MMCs through the incorporation of graphene. For instance, Xiong et al. [24] introduced graphene in Cu by the reduction of reduced graphene oxide through sintering. Similarly, ball milling, molecular-level synthesis, spark plasma sintering, and epitaxial growth have been used to improve the strength of composites using graphene as a reinforcement [13, 25-28]. However, each of the these strengthening techniques has some limitations. For instance, ball milling and molecular-level mixing may allow a uniform dispersion of the reinforcement material but may impart structural defects due to the shear stress and the contamination during the fabrication process [29]. A well-ordered/-aligned, uniformly dispersed, and continuous graphene network is essential to attain the best reinforcement results [13]. Kawk et al. [30] introduced a simple, economically efficient two-step process with the potential to deliver better-quality products with uniformly dispersed and continuous graphene networks.

The two-step process involves the compaction of a metallic powder followed by CVD. In this study, we fabricated a 3Di-hBN-Cu-Ni composite using a similar simple two-step process. Various characterization techniques were employed to confirm the formation of 3Di-hBN surrounding the grains of the Cu-Ni alloys. Cu-Ni-based alloys have been applied in various industries, such as shipbuilding, construction, and processing, owing to their high mechanical strength and corrosion resistance at elevated temperatures. The 3Di-hBN-Cu-Ni composite is expected to deliver better corrosion, mechanical, and wear characteristics than the Cu-Ni alloy. Moreover, the 3Di-hBN layer, a foam-like 3D porous structure, was separated from the 3Di-hBN-Cu-Ni composite. Such a 3Di-hBN material can be applied in the fields of biomedicine, electronics, and energy storage [31-33].

## 2 Experimental Procedures

### ***2.1 Fabrication of 3Di-hBN-Cu-Ni Composite***

Cu powder (99.5 % purity) with spheroidal particles of size 14–25 µm and Ni powder (>99.5% purity) with spheroidal particles of size ~1 µm were purchased from Sigma-Aldrich and used after heat treatment (200°C for 2 h in an $H_2$ environment) to remove any moisture or oxide contents. The chemical compositions of Cu and Ni powders are provided in Table 1. Cu and Ni powders (70 wt.% Cu, 30 wt.% Ni) were mixed manually using mortar with care not to change the particle size distribution and the mixture was compacted in a mold using a double-action oil hydraulic press at the compaction pressures of 60, 110, 220, 280, 335, and 390 MPa. The exertion of high pressure on the spheroidal particles caused mechanical cold locking among the particles, thus forming a compact disc with the approximate diameter and thickness of 15 mm and 1.2 mm, respectively. As shown on the fracture surface of the cross section of the compact disc in Figure 1, relatively large Cu particles produced mechanical interlocking owing to their deformation, and Ni particles filled the gaps between the Cu particles. The discs were then placed in a quartz glass tube furnace with a tube diameter of 23 mm for metal-organic CVD (MOCVD). The compaction pressure and sintering time were varied to determine the optimum conditions for the fabrication of 3D-hBN in the 3Di-hBN-Cu-Ni composite.

Table 1. Chemical compositions of Cu and Ni powders

| Material | Purity | Trace metals in ppm |
|---|---|---|
| Cu Powder | > 99.5% | Fe 80.0, Na 9.42, Mn 7.6, Mg 4.69, Al 4.4, and B 1.98 |
| Ni Powder | > 99.5% | Ag 1.3, Al17.9, Ba 0.8 Ca 27.9, Cr 3.6, Cu 305.2, Fe 383.7 Mg 2.0, Mn 2.6, Na 11.1, Pd 8.0, Ti 144.5, and V 25.4 |

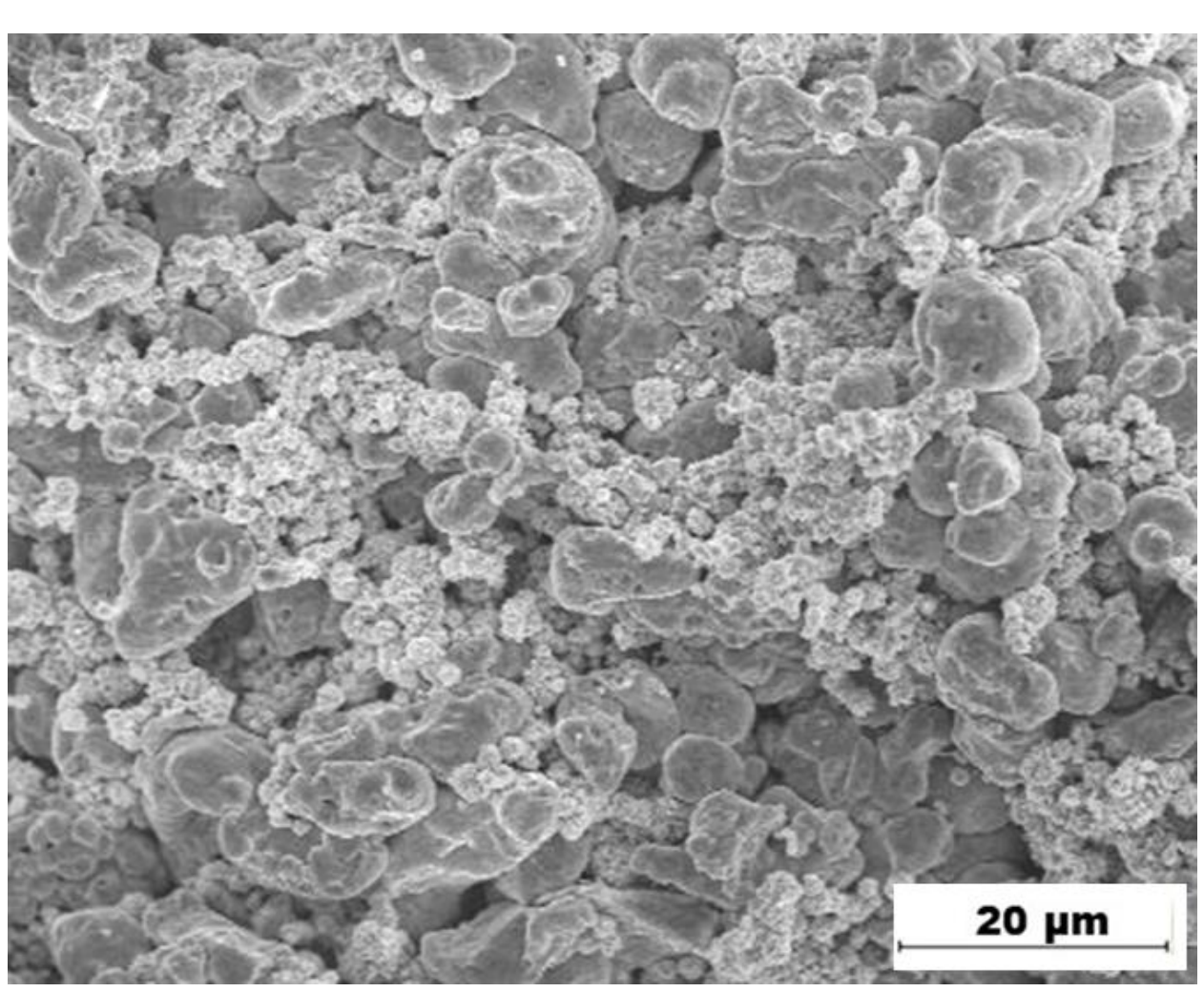

**Fig. 1**. Cu and Ni particles shown at the fracture surface of the disc after compaction

Figure 2 (a) shows a schematic of the system used to fabricate the 3Di-hBN-Cu-Ni composites. Initially, the system was flushed with argon at least three times to remove air from the MOCVD tube. The furnace temperature was increased to 400°C at a rate of 16.6°C/min and then maintained constant in a hydrogen environment at 330 Torr for 1 h for deoxidation. Subsequently, the temperature was raised to 1000°C at the same rate and maintained constant for 15 or 30 min. Finally, MOCVD was performed for 15 min at 450 Torr using heated decaborane ($B_{10}H_{14}$) as the boron source and ammonia ($NH_3$) as the nitrogen source. Decaborane was the preferred boron precursor because of its (i) easy handling, (ii) commercial availability, and (iii) stability, which minimized the formation of undesired side products at elevated temperatures that could potentially decrease hBN yield [34]. Decaborane is a crystalline solid with a melting temperature of 98–100°C and its vapor pressure can be easily controlled by varying the temperature from room temperature to 100°C. At approximately 100°C, the vapors produced upon evaporation can be transported into the MOCVD growth zone by an inert carrier gas (Ar) at a flow rate of 1 sccm. Ammonia gas was introduced as a nitrogen source in the MOCVD reaction zone at a flow rate of 2 sccm. At 1000°C, ammonia and decaborane dissociated into nitrogen and boron atoms, respectively.

$$NH_{3(g)} \rightarrow [N] + \frac{3}{2}H_{2(g)}, \qquad (1)$$

$$B_{10}H_{14(g)} \rightarrow 10[B] + 7H_{2(g)} \qquad (2)$$

The entire process (heating, sintering, and MOCVD) was conducted at a hydrogen flow rate of 10 sccm. The 3Di-hBN-Cu-Ni composite fabricated using a simple two-step process is shown in Figure 2 (b).

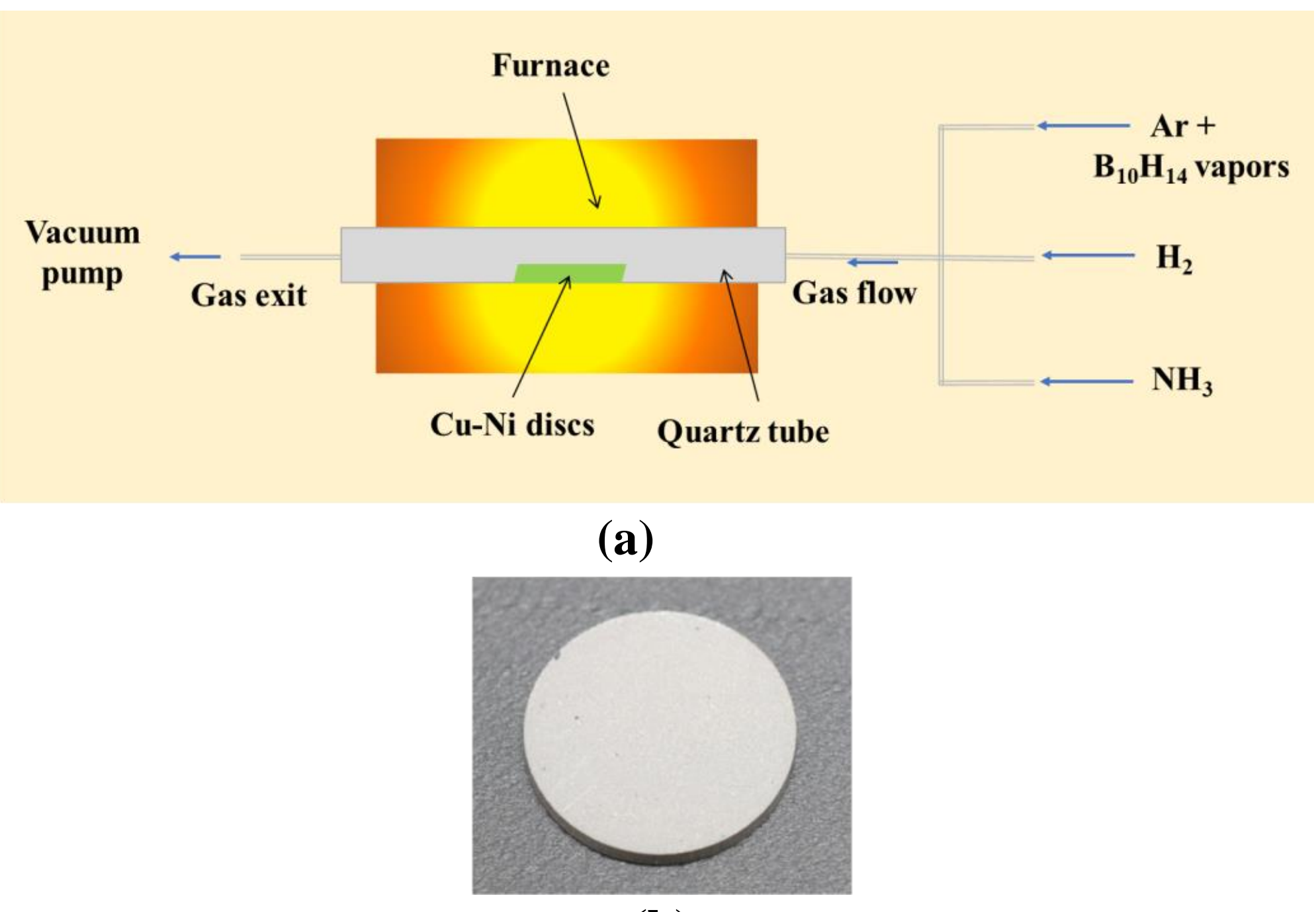

**Fig. 2** (a) Schematic illustration of the fabrication of 3Di-hBN-Cu-Ni composite and (b) disc-shaped 3Di-hBN-Cu-Ni composite fabricated using a simple two-step process

### *2.2 Characterization*

Optical microscopy (OM) and scanning electron microscopy (SEM) investigations of the 3Di-hBN-Cu-Ni composite samples were conducted after the sample was cut in half and the cutting surface was polished with emery papers of size down to 4000 grit. Finally, the polished surface was etched at room temperature using a mixed solution of 1 M $FeCl_3$ and 0.1 M HCl to reveal the microstructure [30]. For transmission electron microscopy (TEM) investigations, the 3Di-hBN-Cu-Ni composite samples were mechanically polished to a thickness of 100 μm and cut into small pieces of 3 mm diameter. Then, Cu-Ni was etched out, leaving only 3Di-hBN foam, which was transferred after thorough cleaning to the TEM grid for investigation. A qualitative phase analysis of the 3Di-hBN-Cu-Ni composite was performed by XRD analysis using Cu $K_\alpha$ radiation with a wavelength of 1.54 Å and a scanning angle of 20°–100°. The density of the 3Di-hBN-Cu-Ni composite samples was measured using the Archimedes immersion technique.

### *2.3 Three-Dimensionally Interconnected hBN*

The 3Di-hBN-Cu-Ni composite samples were cut into small pieces, polished, and placed in an etchant for a sufficient duration to etch out Cu-Ni completely such that only 3Di-hBN remained. Then, the foam-like 3Di-hBN samples were removed and washed several times with deionized water. To obtain a stable 3D structure of 3Di-hBN, the freeze-drying method was used to ensure that there was no effect of liquid capillary force and that 3Di-hBN did not structurally collapse [35].

## 3 Results and Discussions

Figure 3 shows a schematic of the processes involved in the synthesis of the 3Di-hBN-Cu-Ni composite. During the sintering, the reduction in volume and the formation of Cu-Ni solid solution occur due to the diffusion of metals under the driving force to reduce the excess surface energy [36, 37]. Consequently, the overall volume of the compact disc is reduced and densification occurs. The formation of 3Di-hBN in the composite is likely to occur in three stages [8]. First, the diffusion of metal occurs to reduce the surface energy resulting in the formation of large particles (consolidation). At the same time, the diffusion of Ni to Cu or vice versa occurs to form a solid solution of Cu-Ni. Next, during the MOCVD process, the dissociation of ammonia and decaborane produces nitrogen and boron atoms that diffuse into the Cu-Ni alloy at 1000°C. Finally, upon cooling, the nitrogen and boron atoms precipitate out and alternately join together to form 2D hBN layer(s) along the interfaces of the Cu-Ni grains [38] resulting in the formation of Cu-Ni composite.

Small pores or voids can form during the sintering as Cu and Ni particles grow to reduce their surface energy. This is probably due to insufficient sintering time or excessive free space among the particles. These pores may also act as catalytic sites for the nucleation and growth of bulk hBN. The small lighter grey areas (indicated by small white loops) in Figure 4 indicate the bulk hBN that accumulated on the pores during the MOCVD process. These pores generated during the sintering process and then filled with bulk hBN during the MOCVD process are undesirable, as they may adversely affect the mechanical, thermal, and wear characteristics of the composite. Therefore, the processing parameters, such as compaction pressure and sintering time, must be varied to determine the optimal conditions for the fabrication of 3Di-hBN-Cu-Ni composites without the formation of bulk hBN.

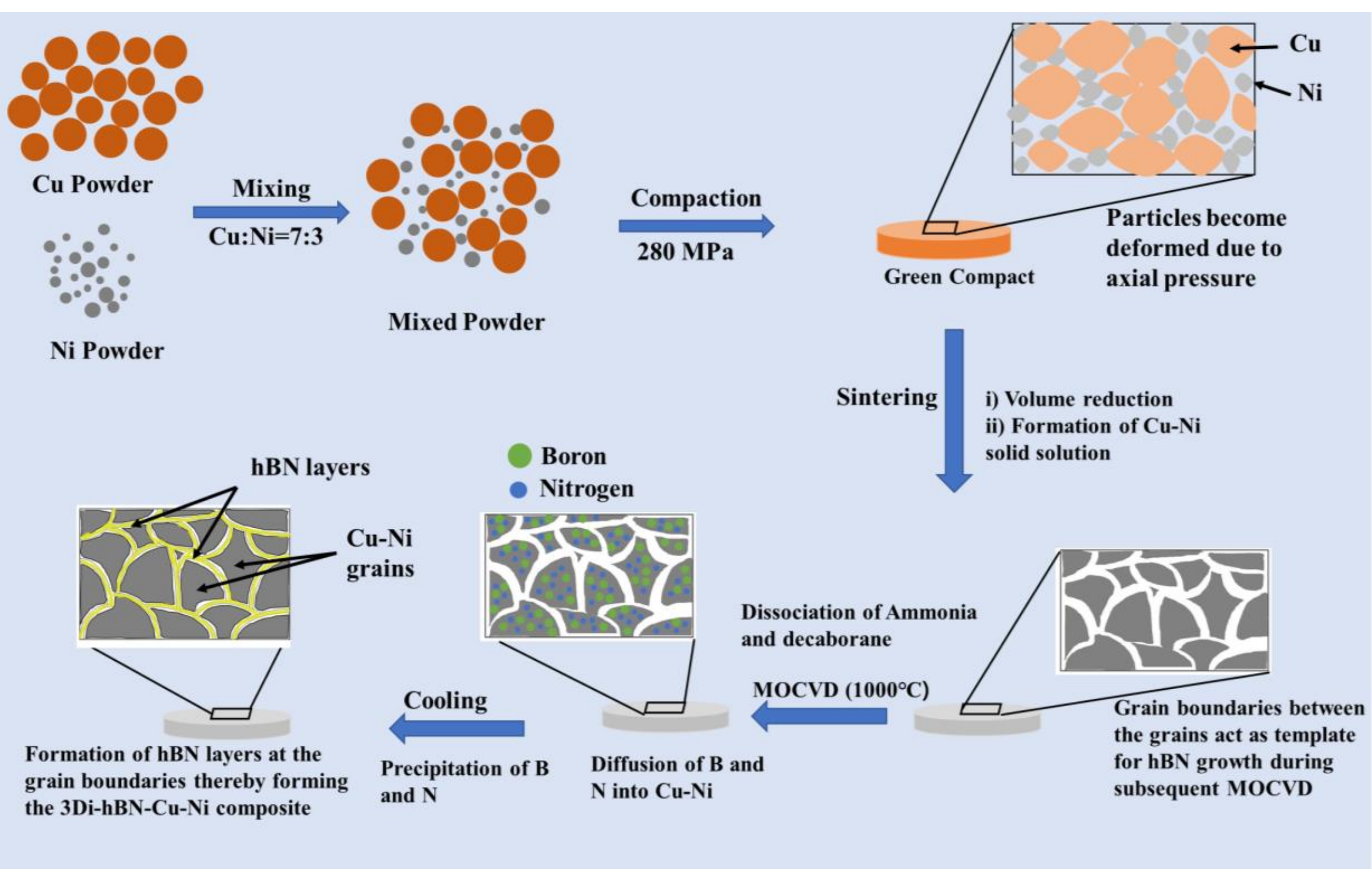

**Fig. 3.** Schematic showing the process of formation of 3Di-hBN-Cu-Ni composite

Figure 5 shows the density of the 3Di-hBN-Cu-Ni composite as a function of the compaction pressure and sintering time. The density of the composite increased with increasing compaction pressure. However, beyond a certain compaction pressure, the density decreased. This trend occurred because at pressures below 280 MPa, the compaction pressure was not enough resulting in low density of compaction; consequently, resulting in lower density of composite having voids or pores. Although these pores were filled with bulk hBN during the subsequent MOCVD process, the overall density of the composite cannot be increased because the density of hBN (2.1 g/cm$^3$) is significantly lower than that of Cu-Ni (~8.9 g/cm$^3$). On the other hand, during compaction at high pressures (>280 MPa), the particles on the surface were pressed with a relatively greater force than those inside the compact disc because of the friction between the particles. Consequently, at pressures exceeding 280 MPa, the surface particles of the compact disc were denser than the inner particles. The inner particles having a longer diffusion distance owing to their lower density resulted in the formation of pores due to insufficient diffusion or a short sintering time, resulting in relatively larger size of pores or shrinkage as indicated by OM images in Figure 6 (b). Hence, the density of the 3Di-hBN-Cu-Ni composite was slightly lower at higher pressures (>280 MPa) as shown in Figure 5. This is also evident from the OM images shown in Figure 4. The white arrows in Figure 4 indicate the bulk hBN present in the microstructure of the 3Di-hBN-Cu-Ni composite. Relatively higher volume fraction of bulk hBN was observed when the compaction pressure is lower or higher than 280 MPa. Furthermore, the density of the 3Di-hBN-Cu-Ni composite also depends on the sintering time as shown in Figure 5. A longer sintering time led to fewer pores (i.e. lower volume fraction of bulk hBN), and consequently more densification occurred.

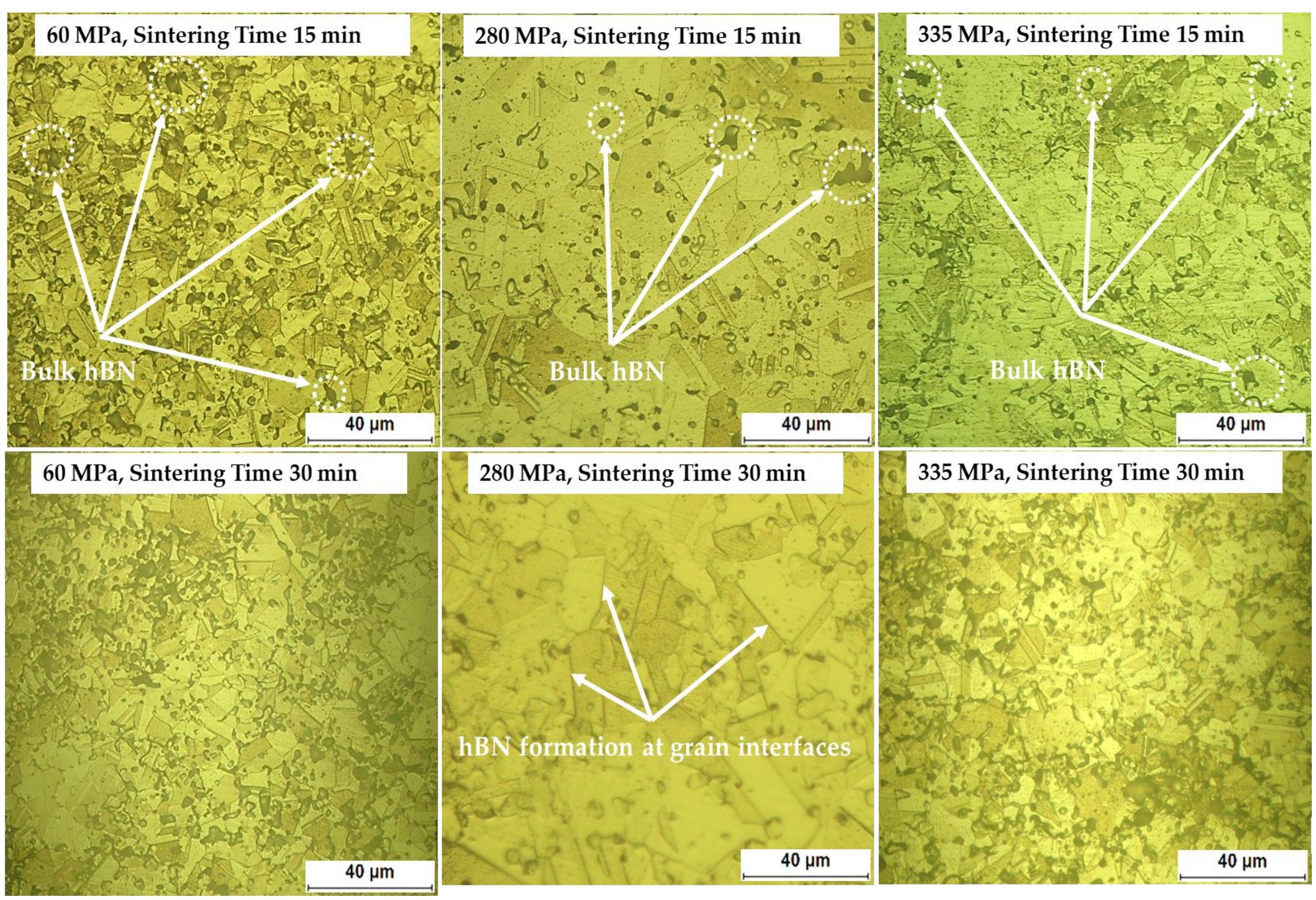

**Fig. 4**. OM images of 3Di-hBN-Cu-Ni composites under various conditions

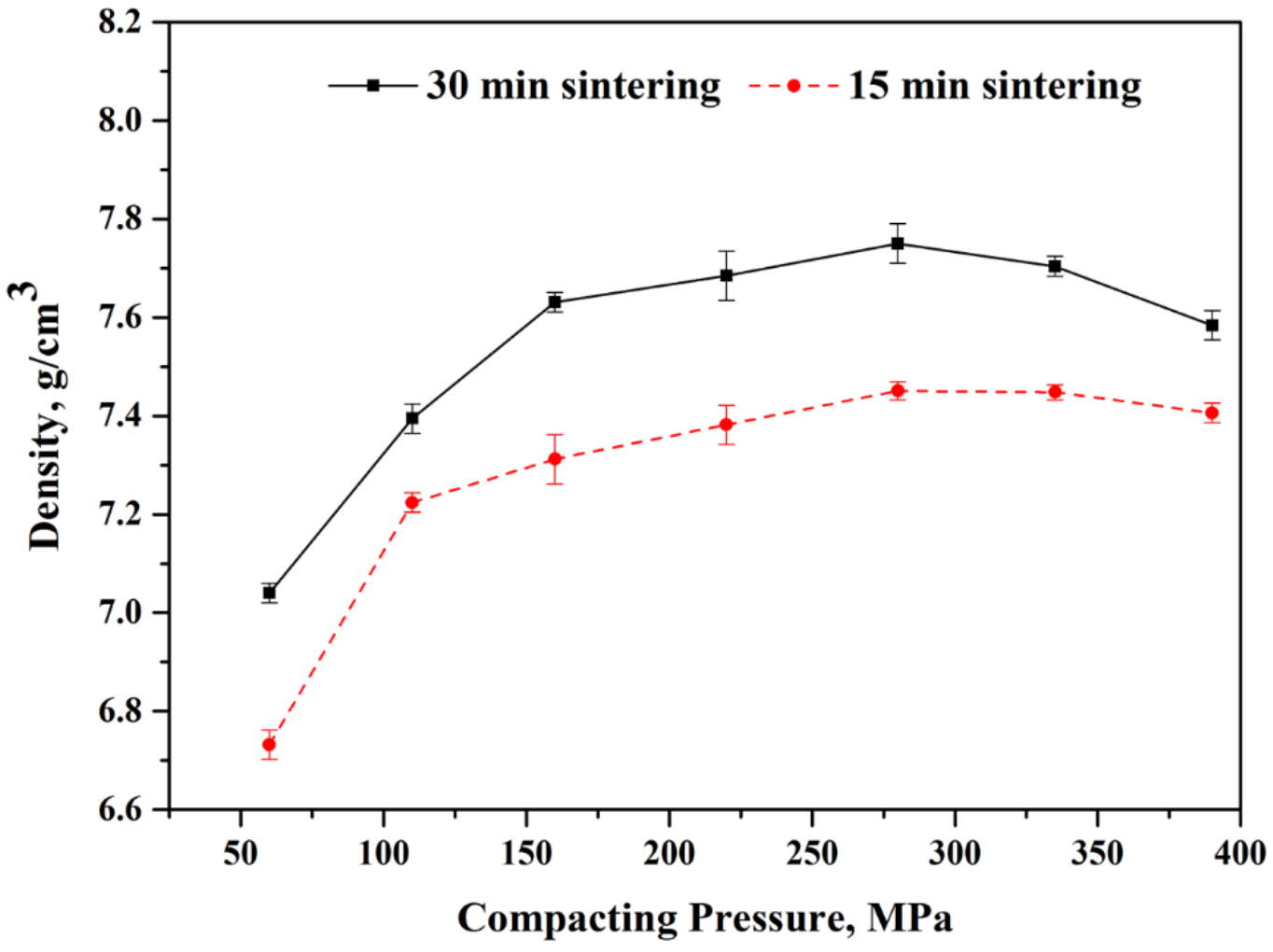

**Fig. 5**. Density of the 3Di-hBN-Cu-Ni composite as a function of compacting pressure and sintering time

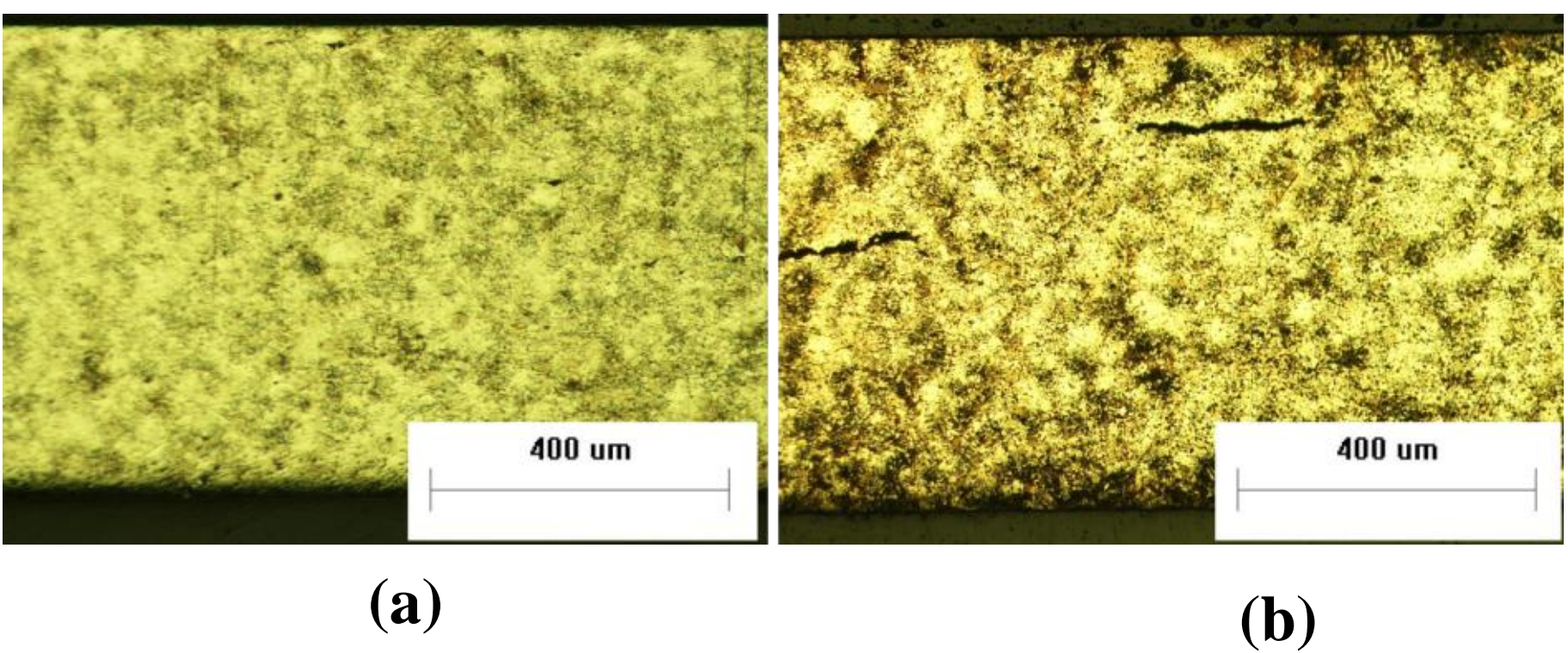

**Fig. 6**. OM images of 3Di hBN-Cu-Ni composites processed at compaction pressures of (a) 280 MPa and (b) 335 MPa

The 3Di-hBN-Cu-Ni-hBN composite, fabricated under the optimized conditions (compaction pressure of 280 MPa and sintering time of 30 min) was examined using SEM. While some of the bulk hBN was removed during the polishing and etching, the SEM image in Figure 7 and the EDS results in Table 2 show the Cu-Ni grains, bulk hBN of less than 5 µm in size, and hBN along the interfaces. The Cu-Ni grains, grain boundaries, and bulk hBN in Figure 7 were analyzed using energy-dispersive X-ray spectroscopy (EDS). Boron and nitrogen were observed (locations (a) and (b) in Figure 7) in excess along with minute amounts of other impurities, such as silicon, carbon, and oxygen, as listed in Table 2. These impurities probably entered the structure during polishing and etching processes. Location (a) in Figure 7 is a pore that was first formed as a consequence of sintering and then filled with bulk hBN during the subsequent MOCVD process. Considering the average size (5 µm) of these sites (location (a)), the presence of bulk hBN was verified through EDS analysis. Further EDS analysis at the grain boundaries (location (b)) revealed that the grain boundaries were also mostly occupied by boron and nitrogen with approximate stochiometric ratio of 1:1. As expected, the Cu-Ni grains (location (c) in Figure 7) comprised Cu and Ni atoms with a ratio of approximately 7 to 3, as shown in Table 2. The solubility of B and N in Cu-Ni alloy at 1000°C is very small (~ ppm) [23] and most of the atoms (B and N) precipitated out during cooling thus forming hBN with B and N having stochiometric ratio of 1:1 at grain boundaries [39-41].

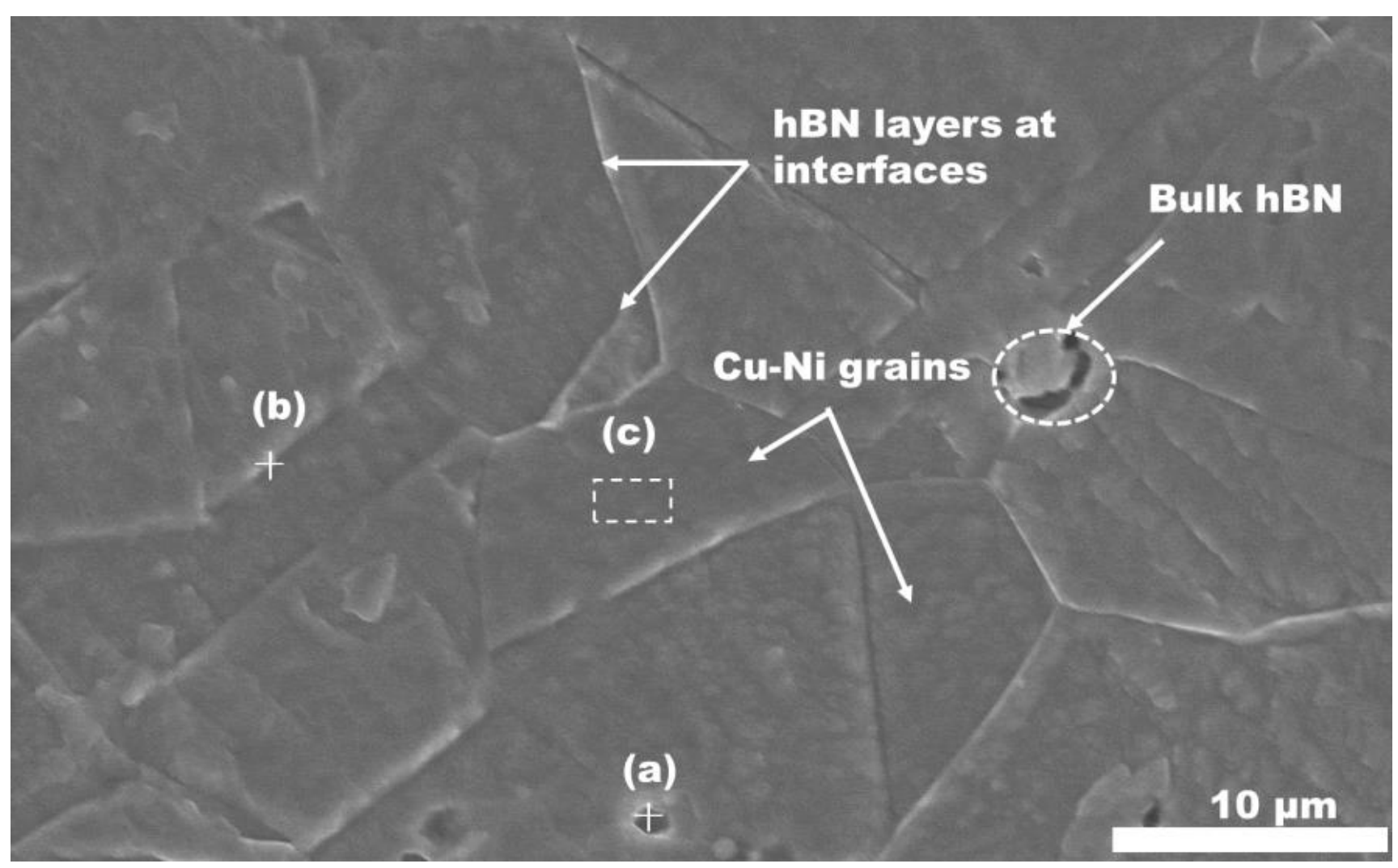

**Fig. 7**. SEM image showing the surface morphology of the 3Di-hBN-Cu-Ni composite fabricated with compaction pressure of 280 MPa and sintering time of 30 min

Table 2. EDS results of the 3Di-hBN-Cu-Ni composite for the microstructure shown in Figure. 7

| **Element** | **Location** | **Cu** | **Ni** | **B** | **N** | **Si** | **C** | **O** |
|---|---|---|---|---|---|---|---|---|
| At. % | (a) in Figure 7 | - | - | 49.32 | 47.45 | 0.76 | 0.87 | 0.68 |
| At. % | (b) in Figure 7 | 20.51 | 8.32 | 34.26 | 34.82 | 0.89 | 0.57 | 0.78 |
| At. % | (c) in Figure 7 | 71.28 | 28.42 | - | - | - | - | - |

The SEM image in Figure 8 shows various hBN layers that interconnect to form a foam-like structure with pockets and channels. The channels are the connected areas between the Cu-Ni grains formed by etching. The average pocket size (10-20 µm) in 3Di-hBN (Figure 8) is approximately equal to the average grain size of the 3Di-hBN-Cu-Ni composite (Figures 5 and 7) indicating that the hBN layers wrapped around the Cu-Ni grains in the 3Di-hBN-Cu-Ni composite.

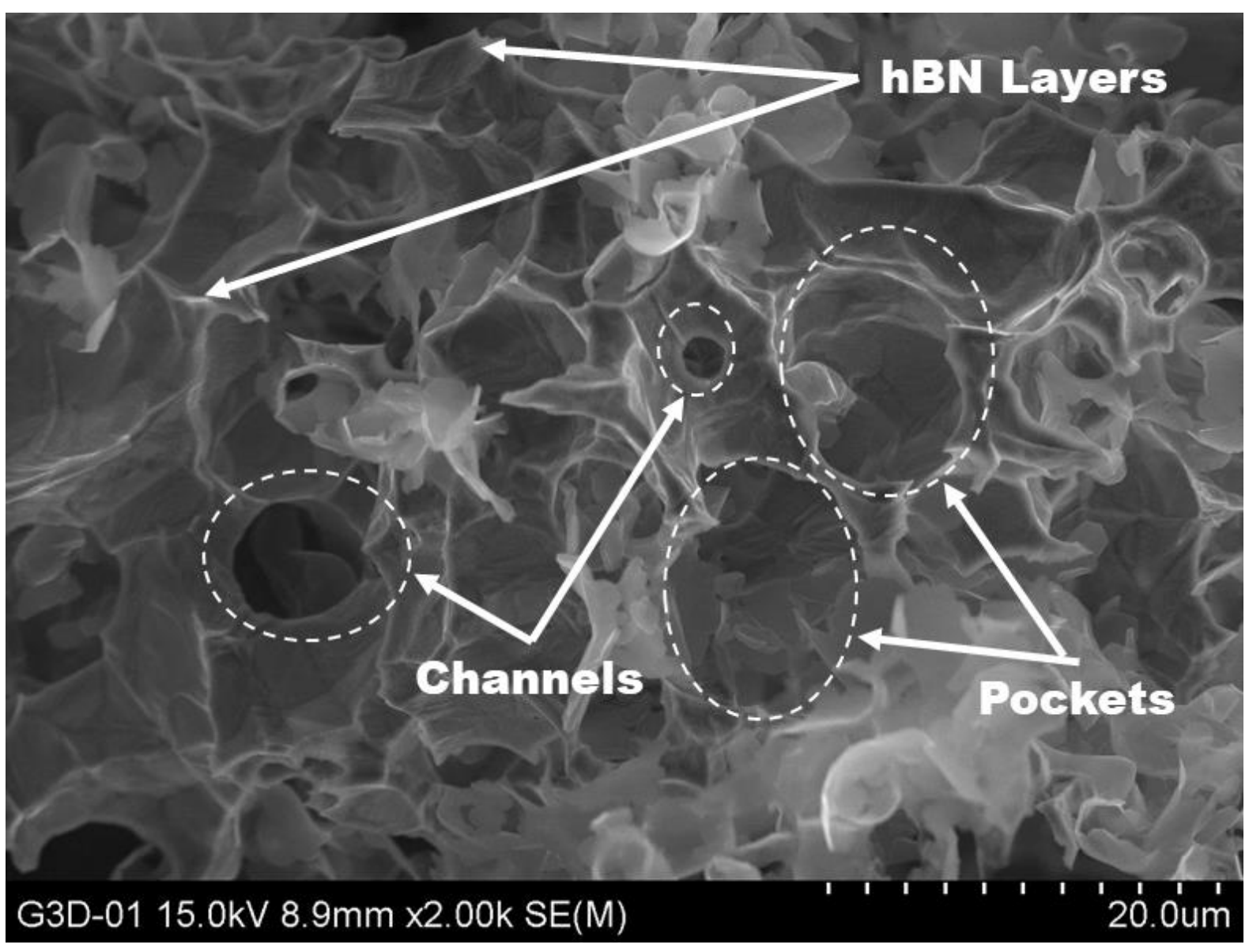

**Fig. 8**. SEM image showing the 3D interconnected network of hBN in 3Di hBN-Cu-Ni composite produced under compaction pressure of 280 MPa and sintering time of 30 min, respectively

The XRD pattern of 3Di-hBN-Cu-Ni composite is shown in Figure 9 (a). This pattern shows the crystalline phase of Cu-Ni solid solution only. Moreover, the elemental distribution map of the 3Di-hBN-Cu-Ni composite shown in Figure 9 (b) shows uniform distribution of Cu and Ni indicating the formation of Cu-Ni solid solution.

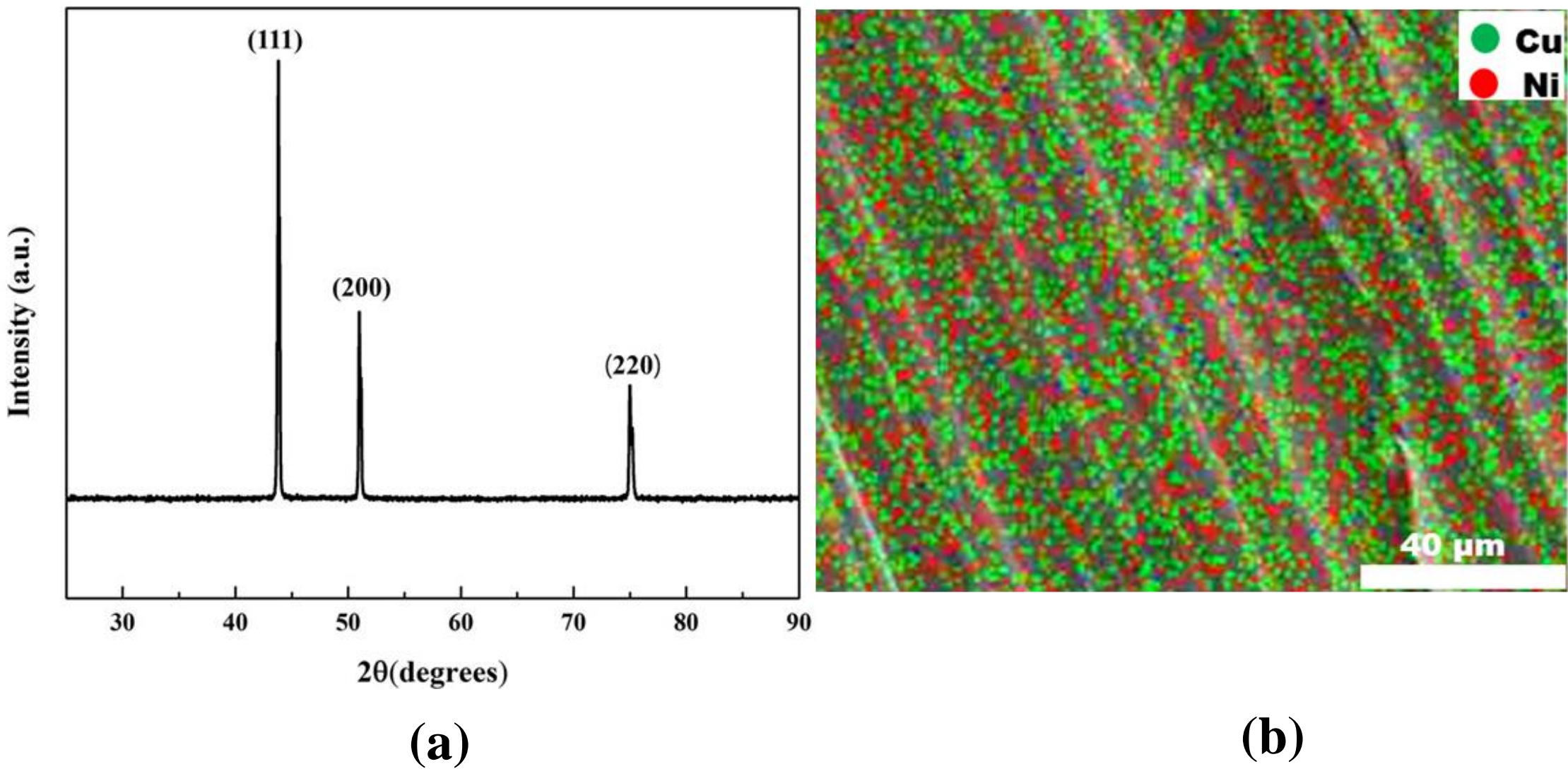

**Fig. 9**. (a) XRD pattern of 3Di-hBN-Cu-Ni composite and (b) Elemental distribution map of 3Di-hBN-Cu-Ni composite

The 3Di-hBN foam was inspected using TEM. The low-magnification bright-field TEM image in Figure 10(a) shows a complex morphology with curvatures and overlapped structures where the 3D layers of hBN (shown in Figure 8) collapsed after their transfer to the TEM grid under the capillary force acting during the drying process. The selected-area electron diffraction pattern of 3Di-hBN (inset in Fig. 10(a)) indicates multiple orientations associated with a couple of layers with different orientations. The high-resolution TEM (HR-TEM) image (Figure10 (b)) reveals 2–6 layers of hBN with an interlayer distance of approximately 0.25 nm (inset in Figure 10(b)), which is attributed to the thickness of a single layer of 2D hBN.

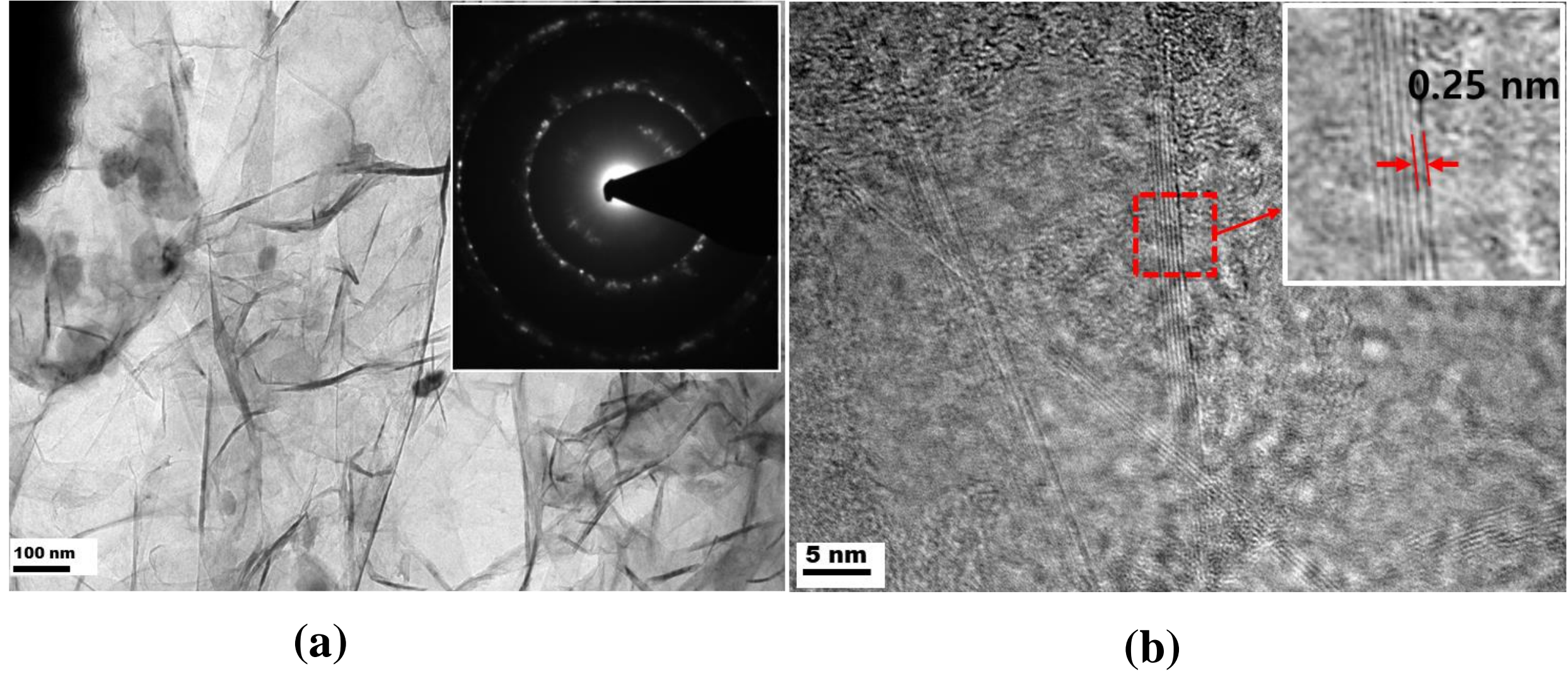

**Fig. 10**. TEM investigation: (a) low-magnification bright-field TEM image of 3Di-hBN; (b) HR-TEM image showing 2–6 layers of hBN; inset shows a distance of 0.25 nm between layers

## 4 Conclusion

3Di-hBN-Cu-Ni composites were synthesized via a simple two-step process of (1) the compaction of Cu and Ni powder mixtures without any additives and (2) MOCVD. The density of the composite was the highest (7.75 g/cm$^3$) when the compaction pressure and sintering time were 280 MPa and 30 min, respectively. OM, SEM, and TEM images indicated that these conditions were optimal for the growth of the interconnected network of hBN in the 3Di-hBN-Cu-Ni composite. SEM investigations and EDS analysis revealed that the grain boundaries were mostly occupied by boron and nitrogen atoms. 3Di-hBN was obtained after etching the Cu-Ni and the average pocket size of the foam was 10–20 µm. The 3Di-hBN-Cu-Ni composite with a density of 7.75 g/cm$^3$ was shown to have a three-dimensional network of 2D hBN. The structural investigation of 3Di-hBN through TEM revealed 2–6 layers with an interlayer

distance of 0.25 nm. This study can be extended further for the characterization of the physical and chemical properties of 3Di-hBN-Cu-Ni composite and 3Di-hBN.

**Acknowledgment**

This study was supported by research funds from Chosun University (2020).